# Study of Warm Electron Injection in Double Gate SONOS by Full Band Monte Carlo Simulation

G. Giusi[(a)], G. Iannaccone[(b)], M. Mohamed [(c)] ,U. Ravaioli[(c)]

[(a)]DEIS, University of Calabria, Via P. Bucci 41C, I-87036 Arcavacata di Rende (CS), Italy

[(b)]DIIEIT, University of Pisa, Via Caruso 16, I-56126 Pisa, Italy(g.iannaccone@iet.unipi.it)

[(c)]University of Illinois at Urbana-Champaign, Urbana, IL 61801 USA

## Abstract

In this paper we investigate *warm* electron injection in a double gate SONOS memory by means of 2D full-band Monte Carlo simulations of the Boltzmann Transport Equation (BTE). Electrons are accelerated in the channel by a drain-to-source voltage $V_{DS}$ smaller than 3 V, so that programming occurs via electrons tunneling through a potential barrier whose height has been effectively reduced by the accumulated kinetic energy. Particle energy distribution at the semiconductor/oxide interface is studied for different bias conditions and different positions along the channel. The gate current is calculated with a continuum-based post-processing method as a function of the particle distribution obtained from Monte Carlo. Simulation results show that the gate current increases by several orders of magnitude with increasing drain bias and warm electron injection can be an interesting option for programming when short channel effects prohibit the application of larger drain bias.

Acknowledgments: Support from the FinFlash IST-NMP Project, from the FIRB Project RBIP06YSJJ and from the RHESSA Project of the Minister of Foreign Affairs is gratefully acknowledged.

## Introduction

Recently, Multi-Gate MOSFET architectures proposed to reduce short channel effects, have been also investigated for non volatile memory applications [1,2]. Multigate architectures keep short channel effects under control for reading bias, but in the case of channel hot electron programming the maximum applicable drain-to-source voltage is limited by punchthrough. In particular, for aggressively scaled devices, $V_{DS}$ cannot be larger than the 3.15 V corresponding to the silicon oxide-silicon potential barrier in the conduction band [1,3]. This means that electrons cannot acquire in the channel sufficient kinetic energy to overcome the potential barrier represented by the gate dielectric, i.e., are not sufficiently "hot". However, experiments show that gate injection for $V_{DS}$ smaller than 3 V is much more efficient than in the case of Fowler-Nordheim programming, meaning that a "warm electron tunneling" mechanism can represent a reasonable option for non-volatile memory programming [1,3].

In this letter we aim to investigate the warm electron tunneling regime and to evaluate its efficiency in SONOS programming. Because data information is stored in the ONO stack through gate tunneling, accurate modeling of the gate current in extremely important for evaluating device performance. In such short devices the transport problem can only be accurately modeled by solving the BTE. However, the gate current is several orders of magnitude smaller than the drain current, and its calculation poses a tremendous challenge to particle-based methods. Attempts to solve the BTE for the gate current problem were made [4,5]. An energy transport model and a Monte Carlo approach were successfully applied to gate current calculations for the case of hot carrier injection [6-9].

In this work we use the Monte Carlo approach to calculate charge density and potential, and we compute the transmission coefficient as a function of energy using the WKB approximation. Size quantization and barrier lowering are neglected. We are interested in particular in the contribution to charge injection due to electrons whose energy is lower than the barrier height (i.e., tunneling electrons).

## Monte Carlo Simulation

Simulations have been performed with the Monte Carlo solver MoCa [10], purposely modified to include the simulation of the gate current with a continuum-based method (as opposed to the particle-based method used to compute particle distributions and transport in the channel). We believe that a full-band 3D Monte Carlo solver including short range particle-particle interaction via the P3M method, already implemented in MoCa [11] would provide a natural and accurate means of computing electron distributions in the channel. In this work we want to examine the concept of warm electron injection, and we prefer to adopt an approximate approach to reduce the computational complexity of the problem by using a 2D full-band version of MoCa, including all the relevant scattering mechanisms: electron-electron interaction is approximately taken into account by self-consistently solving the Poisson equation on a fine grid, without explicitly introducing electron-electron scattering mechanisms such as those proposed in Ref. [12]. In the proposed method the gate current is calculated with a post-process approach by extracting the particle distribution in position and energy from the Monte Carlo solver. Let us refer to the double-gate SONOS structure shown in Fig. 1 where $y$ is the direction of tunneling  Our model assumes that the total carrier energy ($E$) and the transversal momentum ($k_x$, $k_z$) are conserved during tunneling, and that the dispersion relation in the oxide is parabolic with isotropic effective mass $m_{ox.}$ The component of the kinetic energy contributing to tunneling is therefore

$$E_y = E_{kin} - \frac{\hbar^2}{2m_{ox}}\left(k_x^2 + k_z^2\right) \tag{1}$$

and the effective barrier height is $\phi_s = B - E_y$, where $B$=3.15 eV is the barrier height of the Si-$SiO_2$ interface. We consider only particles that are at the Si-$SiO_2$ interface and have a positive velocity ($v_y$) in the tunneling direction. For each particle, we can calculate $E_y$ from (1) and compute the quantity $nv(x,z,E_y) \equiv \langle v_y n \rangle$. Refs. [13] and [14] have shown that with proper barrier parameters the I-V characteristics of thin gate stacks can be reproduced with reasonable accuracy of several orders of magnitude without taking into account barrier lowering and with the WKB approximation. We therefore compute the transmission coefficient $T=T(E_y)$ as in Ref [13]. The tunneling current density can be calculated using the formula

$$J_G(x,z) = q\int \langle v_y n(x,z,E_y) \rangle T(E_y) dE_y \qquad (2)$$

An alternative approach consists of assuming that only the total energy is conserved during tunneling, without any assumption on oxide bands. In this case $E_y$ can be calculated by

$$E_y = E_{kin}(k_x,k_y,k_z) - E_{\min}(k_x,k_z) \qquad (3)$$

where $E_{min}$ is the minimum energy that can be obtained by conserving the transversal momentum $k_x$ and $k_z$. In other words, $E_{\min} = \min_{ky}\{E_{kin}(k_x,k_y,k_z)\}$ , where $k_y$ can vary in the whole Brillouin zone. This solution is identical to the previous one only if the band structure in the silicon and in the oxide are parabolic, and the masses along $k_x$ and $k_z$ are identical. The advantage of this second solution is that we do not have to take into account the band index and the particular conduction band minimum, at the price of some computational cost for obtaining $E_{min}$.

## Simulations

The simulated structure is a n-channel DG SONOS memory with a 50 nm channel length and a 4/5/5 nm ONO stack (Fig.1). In Fig. 2 we show the distribution of the kinetic energy $E_{kin}$ for electrons at the silicon-oxide interface in various positions along the channel. As can be seen, only at the source (x=0 nm) electrons obey a Maxwell-Boltzmann distribution. The distribution progressively differs from a displaced equilibrium distribution when approaching the drain side, and it has a maximum at a kinetic energy value very close to the potential energy drop with respect to x=0 nm, which would correspond to ballistic electron transport. In addition, the distribution becomes more and more asymmetric with respect to the maximum and in the drain region it is rather flat for energies up to the maximum, with a thermal equilibrium tail for larger kinetic energies corresponding to the lattice temperature. Since transport is partially ballistic, for a drain voltage $V_D$, a significant portion of electrons injected from the source have in the vicinity of the drain a kinetic energy $eV_D$ so that they see a barrier towards the gate of approximate height $B - eV_D$. Such effective barrier lowering significantly increases local tunneling close to the drain.

Figure 3 shows the gate current density $J_G$ as a function of the longitudinal position for different drain bias. Gate current density was calculated by (2), and the transmission coefficient was calculated by assuming parabolic oxide bands (Eq. 1). No significant difference in the gate current has been observed by considering effective bands (Eq. 3). As one can see, the drain voltage significantly increases gate tunneling, which also increases by several orders of magnitude between source (x = 0 nm) and drain (x = 50 nm). It is very important to underline that injection is important also for $eV_D$ lower than the barrier height (3.15 eV) so that gate programming can be obtained also by “warm electrons”.

One should also keep in mind that gate current densities may change significantly in time, due to nitride charging. In our calculations this effect is not considered and the nitride layer is assumed to be neutral, as it is in the initial phase of the program operation. Still, our approach should provide significant information on the injection mechanisms.

Figure 4 shows the “programming efficiency” defined in this case as the ratio of the maximum gate current density along the channel to the drain current for different program bias conditions and for different gate lengths (it is an inverse length). In the warm electron injection regime ($V_{DS}$ < 3V) the programming efficiency exponentially increases with increasing $V_{DS}$, whereas it almost saturates for $V_{DS}$ > 3 V. This means that since short channel effects in nanoscale SONOS memories pose an upper limit to the maximum applicable drain bias during programming, warm electron injection can provide an interesting option for achieving fast programming speed ,as confirmed by experimental results [1,3,15].

## Conclusions

In this work we have investigated through 2D full-band Monte Carlo simulations the gate current injection in a double-gate SONOS memory programmed with warm electrons, i.e. with $V_{DS}$ smaller than the silicon oxide-silicon barrier height. Particle energy distribution at the semiconductor/oxide interface are extracted at different bias conditions and different positions along the channel. Gate current is calculated during a post-processing phase as a function of the particle distribution, neglecting size quantization. Simulation results show that injection is effective also for low drain biases because of the very strong dependence of gate current on $V_{DS}$. Warm electron injection could be an interesting option for very short devices for which punchthrough does not allow to apply larger $V_{DS}$.

## Figures

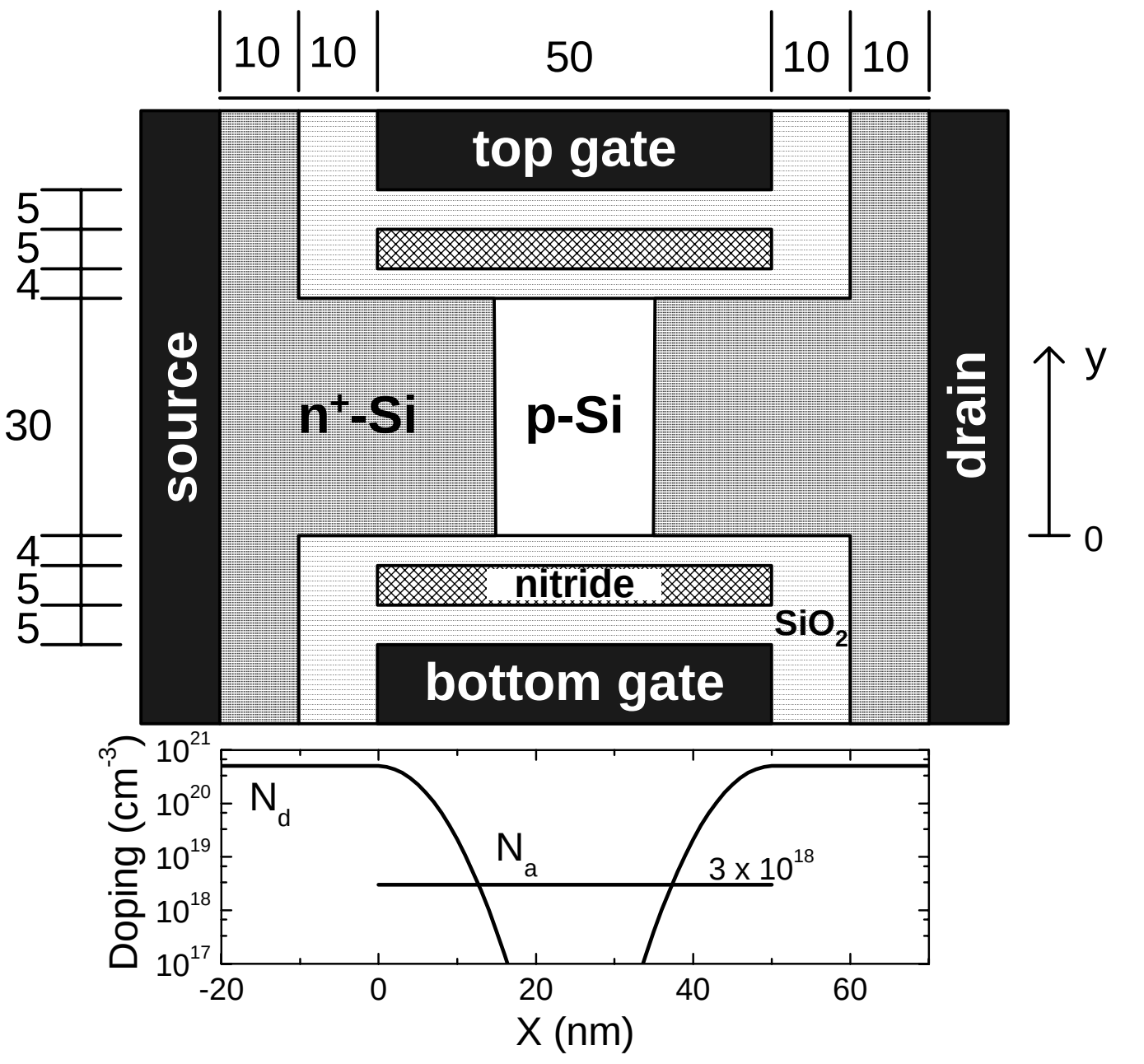

Figure 1

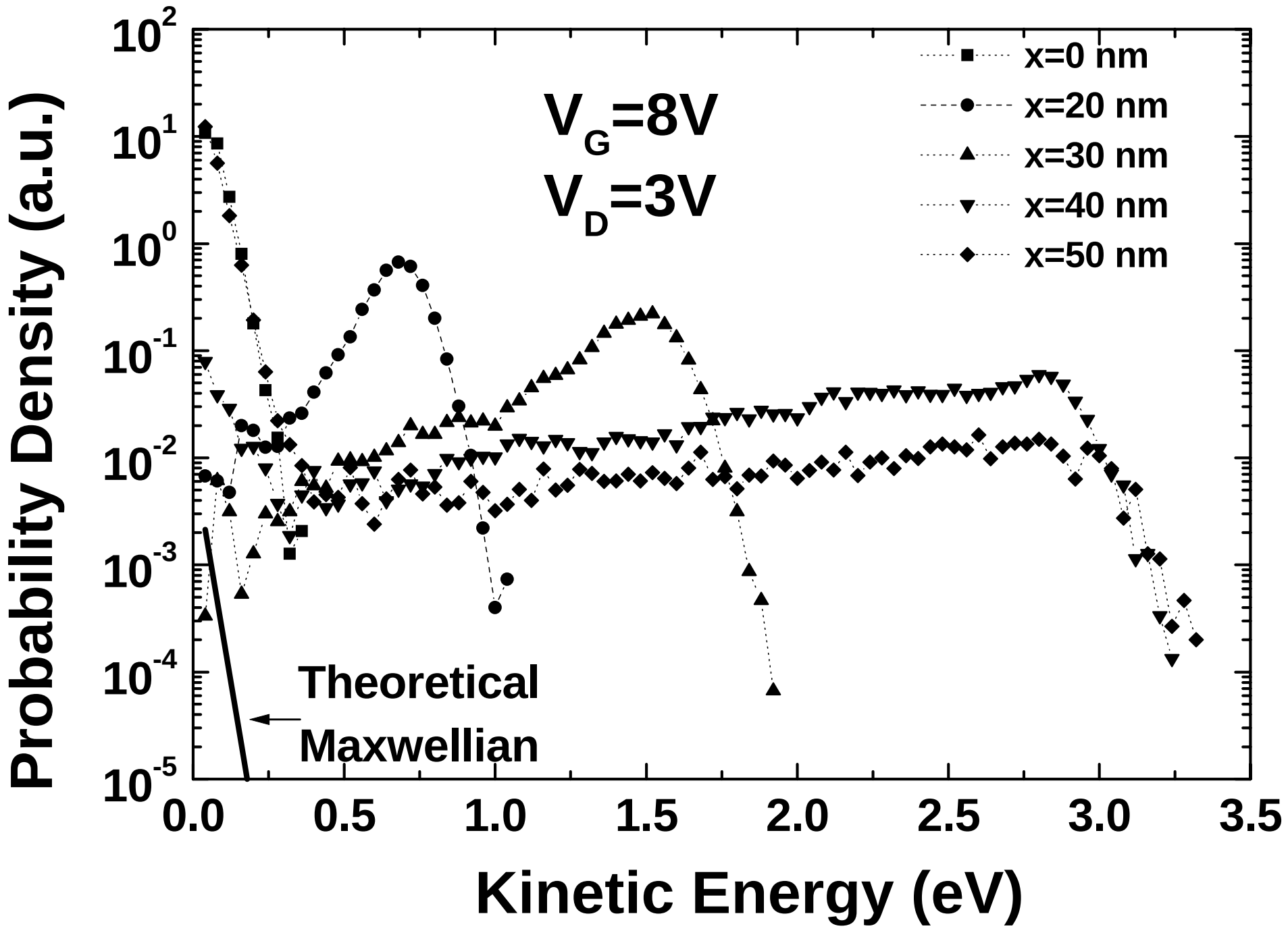

Figure 2

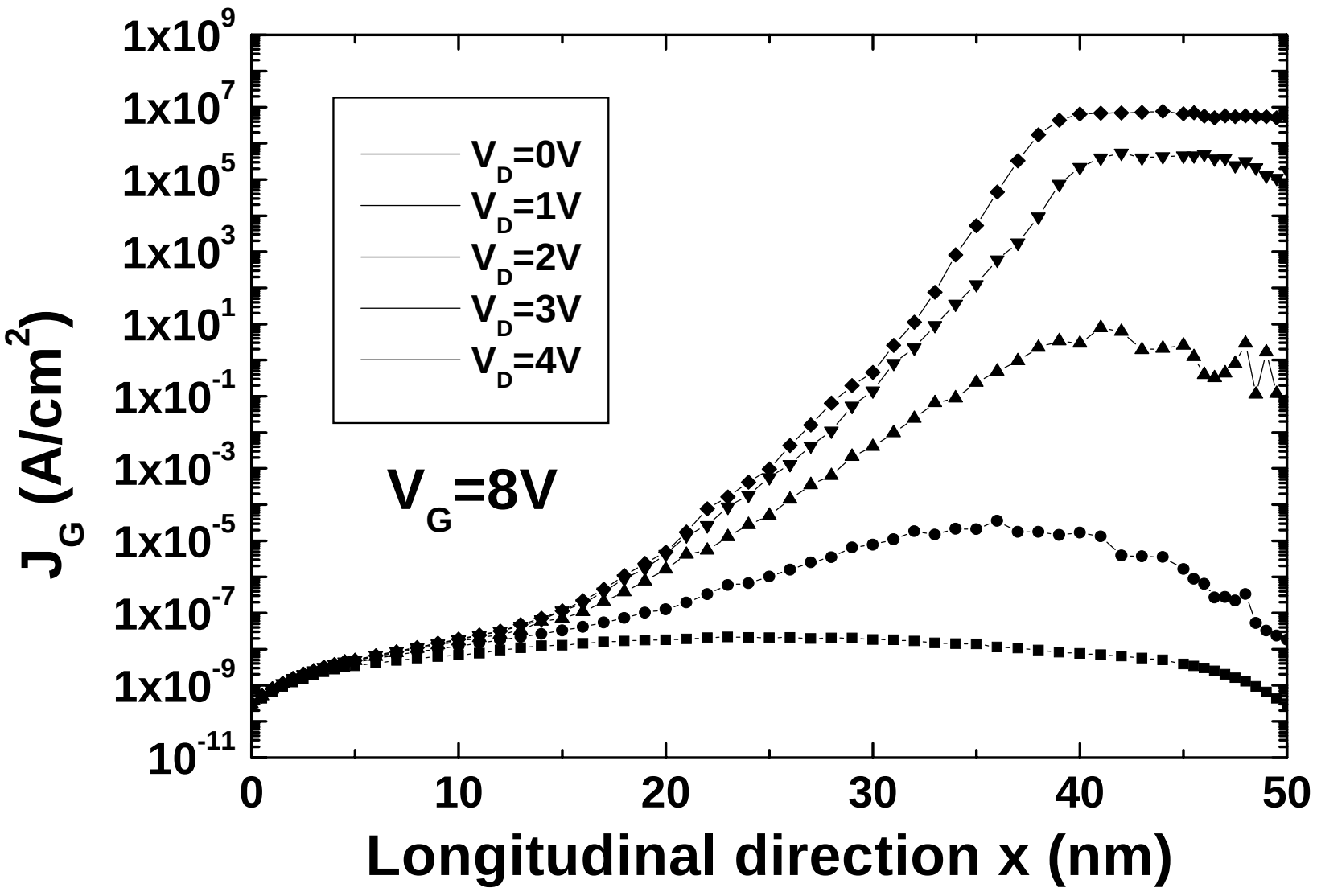

Figure 3

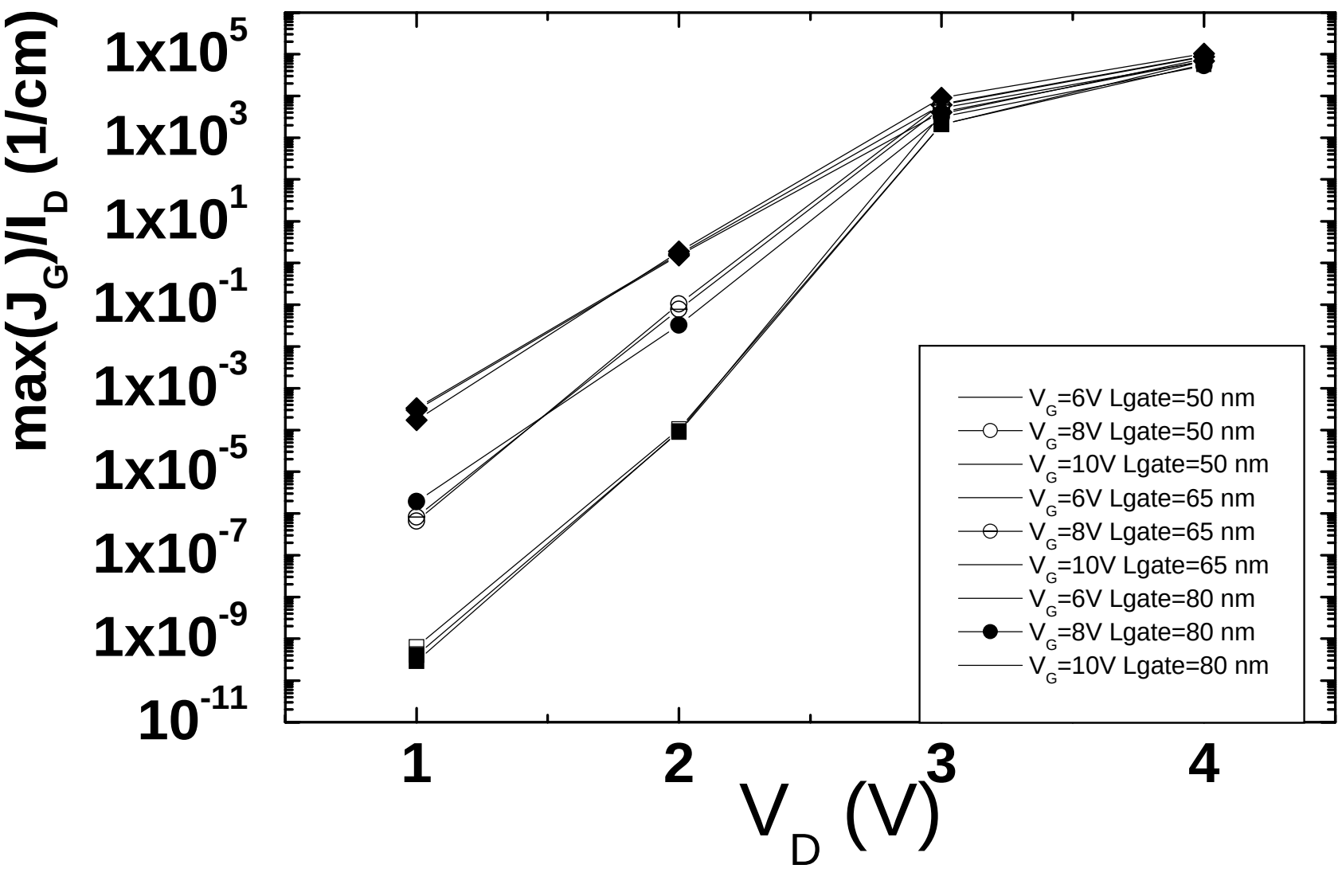

Figure 4

## Figure Captions

Fig. 1

The simulated structure is a double gate SONOS memory with a 4 nm tunnel oxide, 5 nm Nitride oxide, 5 nm top oxide, a fin width of 30 nm and a gate length of 50 nm. The acceptor Fin Doping is $3 \cdot 10^{18}$ $cm^{-3}$, while the source/drain doping extends under the gate for 15 nm from each side. $y$ is the direction of tunneling (perpendicular to the Si-$SiO_2$ interface). The interface is at $y$=0, $y$>0 is silicon, $y$<0 is oxide.

Fig. 2

The distribution of the kinetic energy $E_{kin}$ for electrons at the silicon-oxide interface in various positions along the channel. The electron distribution obeys a Maxwell-Boltzmann law only at the source (x=0 nm) and it progressively departs from an equilibrium distribution as the drain is approached.

Fig. 3

The gate current density $J_G$ as a function of the longitudinal position for different drain bias. The drain voltage significantly increases gate tunneling, which also increases of several orders of magnitude from the source (x=0 nm) to the drain (x=50 nm).

Fig. 4

The programming efficiency defined as the ratio of the maximum gate current density in the channel to the drain current. In the warm electron injection regime ($V_D$ < 3V) the programming efficiency strongly increases with increasing VDS, whereas is almost saturates for $V_D$ > 3 V.